# 2D Semiconductors Superlattices as Hyperbolic Materials

Ilana Kats[1], Tomer Eini[1], and Itai Epstein[1,2,3]

[1] School of Electrical Engineering, Faculty of Engineering, Tel Aviv University, Tel Aviv 6997801, Israel
[2] Center for Light-Matter Interaction, Tel Aviv University, Tel Aviv 6997801, Israel
[3] QuanTAU, Quantum Science and Technology center, Tel Aviv University, Tel Aviv 6997801, Israel

Hyperbolic materials are natural or engineered artificial structures that provide means to manipulate and control electromagnetic radiation, leading to a variety of strong light-matter interactions at the nanoscale. In this work, we explore the physical properties of the optical response of 2D semiconductor-based superlattices, which are engineered with atomic precision and composed of alternating 2D semiconductor monolayers and hexagonal-boron-nitride. We find that such superlattices exhibit a robust hyperbolic response at the visible to near-infrared spectrum, and with dimensions that are an order of magnitude smaller compared to conventional metal-based hyperbolic metamaterials, down to three active monolayers. By employing both analytical and numerical studies, we show that by varying the superlattice configuration we can control and manipulate the nature of the hyperbolic response. Finaly, we propose an optimal structure that will enable the experimental observation of this hyperbolic response in 2D semiconductors superlattices. Such highly compact hyperbolic materials of atomic precision could open the way for deep-subwavelength optoelectronic devices with extremely small footprint.

Hyperbolic materials are anisotropic materials in which the permittivity varies in different axes and possesses opposite signs: $Re\{\varepsilon_\parallel\} \cdot Re\{\varepsilon_\perp\} < 0$, where $\varepsilon_\perp$ and $\varepsilon_\parallel$ are the in-plane and out-of-plane permittivities relative to the optical axis, respectively. These materials exhibit the following dispersion relation for TM plane waves[1–3]:

$$\frac{q^2}{\varepsilon_\parallel} + \frac{\beta^2}{\varepsilon_\perp} = k_0^2 \quad (1)$$

where $q$ and $\beta$ are the transverse $x$ component and the longitudinal $z$ component of the wave-vector respectively, and $k_0$ is the free-space wave-vector. The obtained hyperboloid isofrequency surface in $k$ space leads to a material response that can support arbitrary large wave-vectors.

Bulk materials with a natural hyperbolic response have not been abundantly found in nature, leading initially to the artificial realizations of hyperbolic metamaterials (HMMs). These are engineered structures, comprising alternating metal and dielectric materials, resulting in an effective permittivity that fulfills the hyperbolicity condition discussed above[1–3]. This behavior can be understood in the 1D lattice case, for example, as stemming from the electron's movement in the in-plane direction of the metal layers, while being limited in the out-of-plane direction due to the presence of the dielectric layers.

Similarly, this behavior can be described by the emergence of surface polaritons at the interface between the dielectric and metal layers[4–9]. Surface polaritons can carry large momentum (wave-vectors) and thus possess shorter wavelength compared to the free-space photon. This property enables light manipulation and confinement below the diffraction limit, which is advantageous for a plethora of light-matter interactions, such as probing of quantum effects, waveguiding, enhancing emission rates, compact opto-electronic devices, ultrasensitive spectroscopy, imaging and sensing[10–19].

Compared to bulk materials, hyperbolic two-dimensional (2D) materials can be naturally found due to the layered structure and intrinsic anisotropy of these materials. Hyperbolic 2D materials, such as hexagonal boron nitride (hBN)[19–22], $\alpha$-MoO$_3$[23–25], $\alpha$-V$_2$O$_5$[26] and WTe$_2$[27], have also been shown to support hyperbolic polaritons in the mid-infrared and terahertz spectrum with extraordinary properties, such as low optical losses, high order modes and structural tunabilty[21,23,24,26].

Recently, several reports have shown hyperbolic and in-plane polaritons based on the excitonic response of 2D semiconductors[28–30]. However, these were either based on a monolayer in-plane hyperbolicity[28], transient response of thick transition metal dichalcogenide (TMD) slabs in the mid-infrared[29], or few-layer TMDs with optical properties that have not yet been verified[30]. Thus, a promising approach to achieve hyperbolic response in atomically thin 2D materials at visible to near-infrared (VIS-NIR) frequencies has not yet been achieved.

In this work, we show that a superlattice structure of atomic precision consisting of alternating layers of monolayer TMDs and hBN, exhibits a robust hyperbolic optical response in the VIS-NIR frequency range, and at dimensions that are an order of magnitude smaller compared to metal-based hyperbolic metamaterials, down to three active monolayers. This hyperbolicity stems from the strong excitonic optical response of monolayer TMDs, which can exhibit negative in-plane permittivities[31]. We study the development of the hyperbolic response in the superlattice by analyzing the dispersion relation, both analytically using the transmission-line-model (TLM) and numerically via the transfer-matrix-method (TMM), and find that by varying the number of TMD monolayers and hBN thickness we can control the nature of the hyperbolic response. Finally, we propose an optimal structure for the experimental observation of the phenomenon, which coincides with recent successful fabrications of similar superlattices[32–38].

We start by examining a superlattice structure consisting of alternating high-quality TMD monolayers of $WS_2$ and hBN layers, as shown in Fig 1(a). We assume the heterostructure is infinite in the $x$-$y$ plane, with the bottom hBN layer modeled as semi-infinite in the $z$ direction. We explore this structure using two methods: analytically using the transmission-line-model (TLM) and numerical simulations based on the transfer-matrix-method (TMM), both which will be further discussed below. For monolayer TMDs, the excitonic response is commonly described by a Lorentzian complex susceptibility, given by[31,39]:

$$\chi_\perp = \chi_{bg} - \frac{c}{\omega_0 d_0} \frac{\gamma_{r,0}}{\omega - \omega_0 + i\left(\frac{\gamma_{nr}}{2} + \gamma_d\right)} \quad (1)$$

where $\chi_{bg}$ is the background susceptibility, $c$ is the speed of light, $\omega_0$ is the exciton energy, $d_0$ is the thickness of the TMD, and $\gamma_{r,0}$, $\gamma_{nr}$ and $\gamma_d$ are the radiative, nonradiative, and pure dephasing decay rate, respectively. In Eq. 1, we use the values of the decay rates as previously obtained from experimental measurements of an hBN-encapsulated $WS_2$ monolayers at cryogenic temperatures[31,38]. We examine first the response of the system at a cryogenic temperature of 10K, and the dependence on temperature will be explored further on in the text.

In this study, TMD monolayers are treated as conducting infinitesimal sheets, and assuming there is only a polarization current in the monolayer, $J_p = \frac{\partial P}{\partial t} = -i\omega P$, caused by the difference in the polarization of the TMD and its surrounding (hBN). Using the connection between polarization and electric field, $P = \varepsilon_0 \chi E$, we can write $J_P = -i\omega\varepsilon_0(\chi_{TMD} - \chi_{hBN})E$. Using Ohm's law, $J = \sigma E$, and multiplication by the TMD monolayer thickness $d_0$ (as Ohm's law relates to volume current), we get that the surface conductivity is $\sigma = -i\omega\varepsilon_0 d_0 (\chi_{TMD} - \chi_{hBN})$ where $\varepsilon_0$ is vacuum permittivity, $\chi_{TMD}$ and $\chi_{hBN}$ are the in-plane susceptibility of TMD and hBN, respectively. We note that hBN is considered as an isotropic dielectric material with a constant permittivity of 3.87, based on the spectral range of interest.

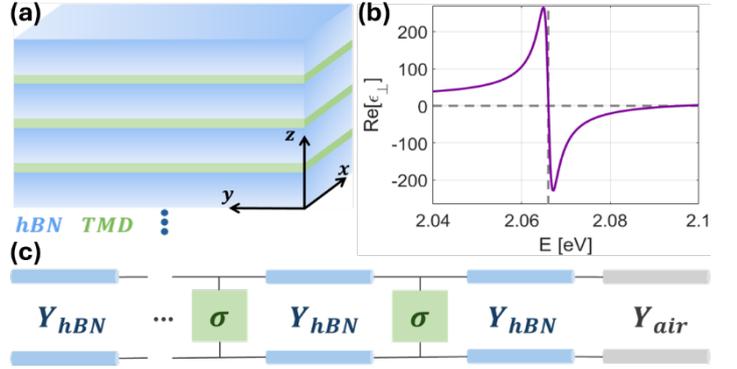

Figure 1: (a) The superlattice structure. Blue layers represent hBN and green represent monolayer TMDs. (b) The real part of $\varepsilon_\perp$ for a $WS_2$ TMD monolayer encapsulated in hBN (purple line). The vertical gray dashed line represents the excitonic resonance energy. (c) The TLM representation of the superlattice.

We use the TMM simulation to explore the optical response of the superlattice, from which we obtain a single matrix describing the reflection, transmission, and absorption coefficients, as well as the dispersion relation from the loss function[40]. In all simulations we use $\gamma_{r,0} = 3.52$ meV, $\chi_{bg} = 17$, $d_0 = 0.618$ nm, $\hbar\omega_0 = 2.067$ eV, $\varepsilon_{hBN} = 3.87$, $\varepsilon_{air} = 1$, and $\hbar\gamma_{nr} = 1.2$ meV, $\hbar\gamma_d = 0.5$ meV[31] unless mentioned otherwise. To calculate the analytical dispersion relation, we employ the TLM, which simplifies the analysis of electromagnetic wave propagation in layered structures by using concepts from electrical circuit analysis[41]. The electric and magnetic fields are analogous to voltage and current in a transmission line and each layer is represented as a segment of a transmission line characterized by its characteristic admittance, $Y_c = \frac{\omega\varepsilon_0\varepsilon_\perp}{\beta}$, (Fig. 1(c)). The TMD monolayer are modeled as a parallel admittance, $Y = \sigma$, as we consider it as a conducting infinitesimal sheet.

Owing to the high confinement of polaritons we can assume $q \gg \sqrt{\varepsilon_\parallel} k_0$, obtaining the relation:

$$\beta \cong iq\sqrt{\frac{\varepsilon_\perp}{\varepsilon_\parallel}} \quad (2)$$

which will be used to calculate the dispersion relation, $q(\omega)$, of the structures at hand. From the structure defined in Fig 1(c), we obtain the following recursive equations:

$$Y_n = Y_{hBN}\frac{Y_{n-1} - iY_{hBN}\tan(\beta d)}{Y_{hBN} - iY_{n-1}\tan(\beta d)} + \sigma \quad (3)$$
$$(n = 1,2,\dots N)$$

$$Y_N = -Y_{hBN} \quad Y_0 = Y_{air} \text{ or } Y_{hBN}$$

where $Y_0$ is the admittance after the first layer from the right, $Y_n$ is the admittance after the nth layer from the right, $Y_{hBN}$ is the characteristic admittance of hBN, $d$ is the thickness of the hBN layers, $\sigma$ is the conductivity of TMD monolayer, and N is the number of TMD monolayers.

First, we treat two simple cases verifying the agreement between the TLM and the TMM. The first case is a single TMD monolayer encapsulated in two semi-infinite hBN layers. The dispersion relation obtained for this case is:

$$q_{single}(\omega) = 2i\frac{\omega\varepsilon_0\varepsilon_{hBN}}{\sigma} \quad (4)$$

and shows a good agreement between the two (Fig. 2(a)).

The second case is a structure containing two TMD monolayers separated by hBN with top and bottom semi-infinite hBN layers. Due to the symmetry of the system, we can separate the problem into even and odd electrical and magnetic modes in the $z$ axis around the center of the structure, denoted as $z = 0$. Owing to the small separation between TMD monolayers we can assume $\frac{\beta d}{2} \ll 1$, and simplify Eq. 3 by $\tan\left(\frac{\beta d}{2}\right) \approx \frac{\beta d}{2}$. From this and Eq. 2 we get $q \ll -\frac{2i}{d}\sqrt{\frac{\varepsilon_\parallel}{\varepsilon_\perp}}$, which implies that the approximation holds for small $q$ compared to $\frac{1}{d}$. The analytic dispersion relation in this case yields:

$$q_1 = -\frac{\omega\varepsilon_0\varepsilon_{hBN}}{\omega\varepsilon_0\varepsilon_{hBN}\frac{d}{2} + \sigma} \quad (5)$$

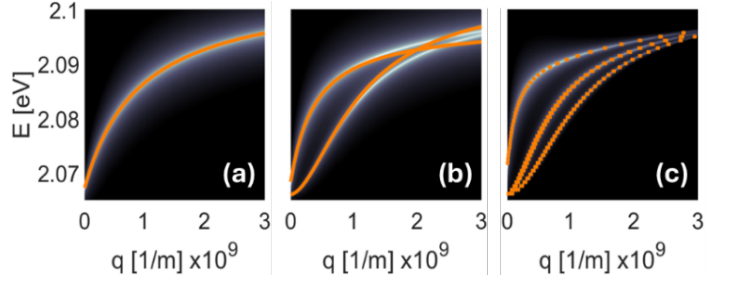

Figure 2: The dispersion relation obtained from the TMM simulation (colormap) and the TLM (orange line) for, (a) a single TMD encapsulated in two semi-finite hBN layers (Eq. 4). (b) a structure containing two TMD monolayers (Eq. 5). (c) a structure of 3 TMD monolayers with the top hBN layer bounded by air. All simulations use $d = 1nm$. In these results the losses have been reduced for a clearer visual comparison between the two methods.

$$q_2 = -i\frac{-\omega\varepsilon_0\varepsilon_{hBN}\frac{d}{2} \pm \sqrt{\left(\omega\varepsilon_0\varepsilon_{hBN}\frac{d}{2}\right)\left(\omega\varepsilon_0\varepsilon_{hBN}\frac{d}{2} - i4\sigma\right)}}{\sigma d}$$

where $q_1$ is the case where the magnetic field is odd and $q_2$ is the case where the electric field is odd, and for $q_2$ a physical solution is obtained only for the expression with the minus sign. In Fig 2(b) we can see that for small values of $q$ the analytic dispersion (orange lines) agrees perfectly with the simulation result (colormap). For larger values of $q$ the solution diverges from the simulation result because the approximation (as stated earlier) doesn't hold anymore.

For a larger number of layers with broken symmetry by including air as the first medium, there is no analytical solution, however, we can still use the simplified TLM representation in a recursive numerical solution. Fig 2(c) provides an example for a structure containing three TMD monolayers, where the top hBN layer is no longer semi-infinite and is instead bounded by air. We can see that the solution of the recursive equations obtained from the TLM agrees well with the simulation for the dispersion relation in this case as well.

Next, we evaluate the number of TMD monolayers needed to observe an effective hyperbolic response of the whole superlattice. Fig 3(a), (b), and (c) show the dispersion relation obtained from the TMM simulations for varying numbers of TMD monolayers, while maintaining a constant hBN thickness. From these figures it is evident that the number of modes increases with the number of TMD monolayers, which agrees well with similar analysis performed on hyperbolic phonon polaritons in hBN[20,42,43], indicating that the superlattice exhibits a hyperbolic response. Further evidence is seen in the reflection of the structure obtained from the TMM simulation, shown in Fig. 3(d). For a very large number of TMD monolayers the reflection peak broadens and reaches large reflectivity values close to 100%, similar to that obtained in the Reststrahlen band of hyperbolic phononic materials[17,19]. In addition, a "dent" at an energy of 2.066 eV (gray dashed line) can be seen to evolve with increasing number of monolayers, which is characteristic of the transverse-optical phonon onset of the Reststrahlen band in hyperbolic phononic materials[17,22].

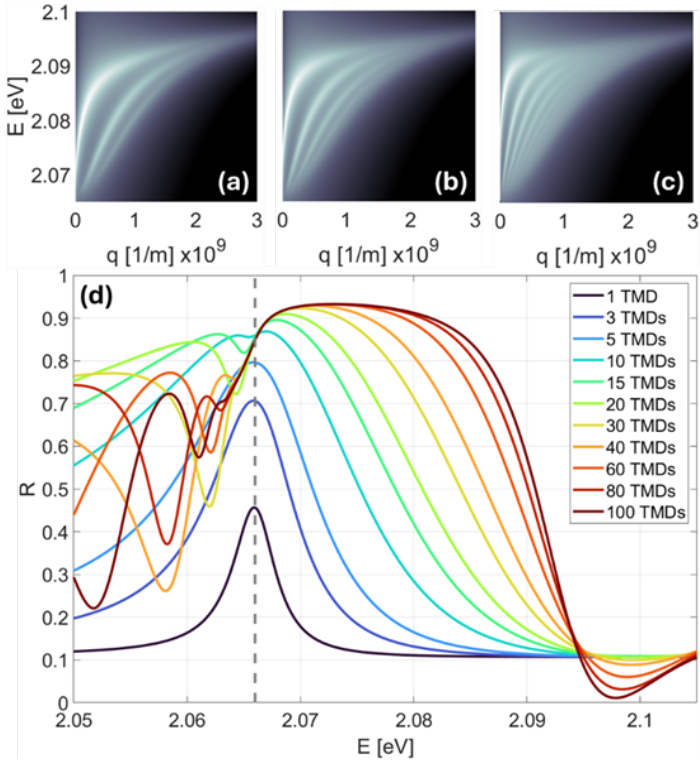

Figure 3: The dispersion relation obtained from the TMM for, (a) a structure with 3 TMD monolayers, (b) a structure with 5 TMD monolayers, and (c) a structure with 10 TMD monolayers. (d) The reflection of the structure for varied number of TMD monolayers (1 to 100). The vertical gray dashed line represents the excitonic resonance. In all ll simulations d = 1nm.

Next, we examine the structure response for varying the thickness of the hBN layers for a constant number of TMD layers. Fig 4(a), (b) and (c) display the dispersion relations obtained from the TMM simulation of a structure containing 3 TMD monolayers and 1nm, 2nm and 10nm thick hBN layers respectively. As the hBN thickness increases, the distance between the dispersion relation branches decreases and eventually converges to a single branch, which is the response of a structure with only one TMD monolayer (orange dots). As the hBN thickness increases and with it the separation between TMD monolayers, the coupling strength decreases until they no longer collectively contribute to the hyperbolic response, effectively behaving as isolated monolayers.

In Fig 4(d), the reflection of the system, as obtained from the TMM simulation, is shown. It can be seen that up to about ~ 20nm, the hyperbolic response becomes stronger, evident by the increased width of the reflection peak and the appearance of a "dent" characteristic of hyperbolic materials. This is due to the increased overall bulk size of the structure that leads to a larger interaction length with the incoming light. We note that the hyperbolic response is clearly observed in this superlattice, which is an order of magnitude thinner compared to conventional metal-based metamaterials[1]. Beyond 20nm the reflection peak gradually narrows, and the "dent" disappears, indicating a weakening of the hyperbolicity due to the weaker interaction between adjacent TMD monolayers.

From the above we can thus unambiguously conclude that the superlattice indeed behaves as a hyperbolic material. Owing to the practical difficulty in fabrication of such multilayer superlattices, we propose a feasible structure for an experimental observation of this phenomenon, consisting of three TMD monolayers separated by 15nm thick hBN layers and placed on a sapphire substrate. Fig. 5(a) and (b) present the real and imaginary part of the in-plane susceptibility, $\chi_\perp$, of a high quality TMD monolayer encapsulated in hBN at various temperatures (10K-300K), based on previous experimental results [31]. It is observed in Fig. 5(b) that lowering the temperature leads to a narrower peak in the imaginary part of $\chi_\perp$ (due to reduced decay rates[31]), while the real part of $\chi_\perp$ gradually becomes more negative for energies above the excitonic resonance (Fig. 5(a)).

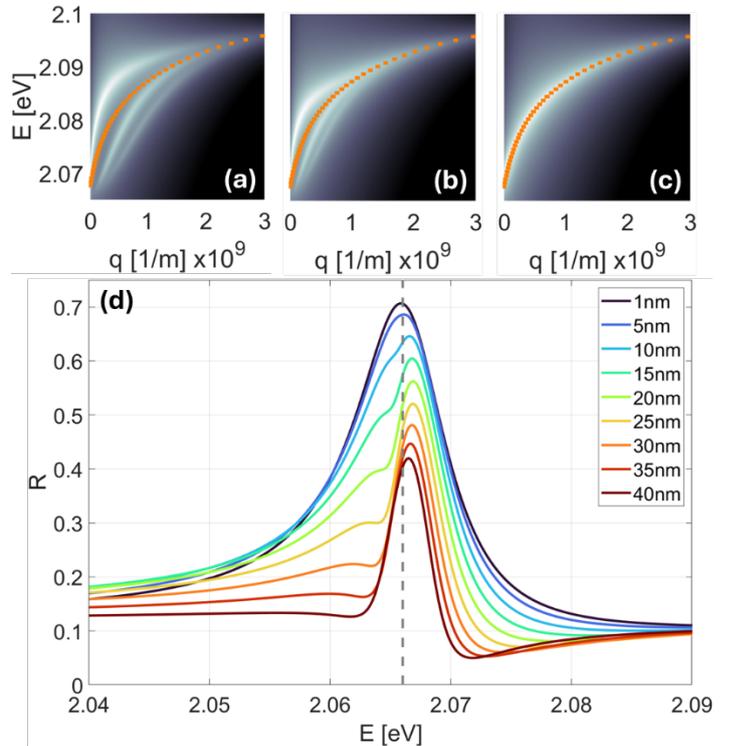

Figure 4: The dispersion relation obtained from the TMM for, (a) superlattice with 1nm hBN layers (colormap) and from the TLM for a single TMD monolayer (orange dots), (b) superlattice with 2nm hBN layers. (c) superlattice with 10nm hBN layers. (d) The reflection of the structure for varied hBN thicknesses (1nm – 40nm). The vertical gray dashed line represents the excitonic resonance energy. All simulations are with 3 TMD monolayers.

Figure 5(c) shows the reflection of the proposed system at different temperatures as obtained from the TMM simulation. As the temperature decreases, the hyperbolic response becomes stronger due to the increasingly negative values of the real part of the TMD's susceptibility, which is not observed for the single monolayer case Fig. 5(d). Comparing Figs. 5(c) and 5(d), we can see that at temperatures below 90K, both structures exhibit a similar reflection shape, indicating no hyperbolic response. However, at lower temperatures, the reflection peak of the superlattice becomes wider than that of the single monolayer structure, with the additional "dent" indicating a hyperbolic behavior.

In conclusion, our theoretical analysis demonstrates that a superlattice of alternating TMD monolayers and hBN layers can exhibit a hyperbolic optical response in the VIS-NIR frequency range. We have shown how varying the number of TMD monolayers and the thickness of hBN layers impacts the strength of the hyperbolic response and proposed a feasible experiment to observe and validate these theoretical predictions. Such compact superlattices could open new opportunities for nanophotonic devices in the VIS-NIR spectrum, such as waveguides, sensors, and optoelectronic devices with controllable properties.

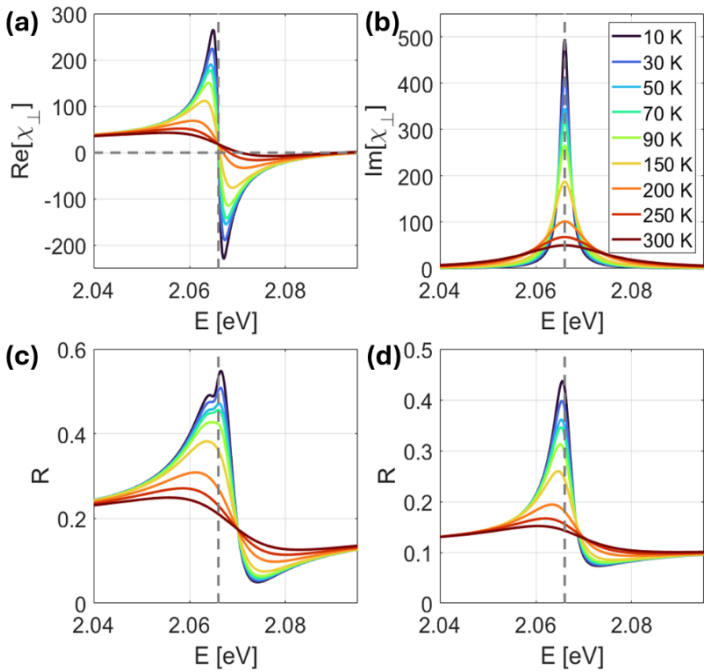

*Figure 5: (a) The real part and (b) imaginary part of the susceptibility of a TMD monolayer encapsulated in hBN for different temperatures (10K – 300K), respectively. (c) The reflection at different temperatures obtained from the TMM for the proposed superlattice experiment. (d) The same as (c) but for a structure containing a single TMD monolayer. The vertical gray dashed line represents the excitonic resonance energy.*

Data availability: The data that supports the findings of this study are available within the article.